\documentclass[apex,twocolumn]{jjap3}
\usepackage{graphicx,newtxtext,newtxmath,float}
\title{Zero-field dynamics stabilized by in-plane shape anisotropy in MgO-based spin-torque oscillators}
\author{Ewa Kowalska$^{1,2}$\thanks{E-mail: e.kowalska@hzdr.de}, Attila K\'{a}kay$^{1}$, Ciar{\'a}n Fowley$^{1}$, Volker Sluka$^{1}$\thanks{Present address: Department of Physics, New York University, 4 Washington Place, N.Y. 10003, New York, USA}, J{\"u}rgen Lindner$^{1}$, J{\"u}rgen Fassbender$^{1,2}$ and Alina M. Deac$^{1}$}
\inst{$^{1}$Helmholtz-Zentrum Dresden - Rossendorf, Institute of Ion Beam Physics and Materials Research, Bautzner Landstrasse 400, 01328 Dresden, Germany\\
$^{2}$Institute of Solid State Physics, TU Dresden, Zellescher Weg 16, 01069 Dresden, Germany}
\abst{Here we demonstrate numerically that shape anisotropy in MgO-based spin-torque nano-oscillators consisting of an out-of-plane magnetized free layer and an in-plane polarizer is necessary to stabilize out-of-plane magnetization precession without the need of external magnetic fields. As the in-plane anisotropy is increased, a gradual tilting of the magnetization towards the in-plane easy direction is introduced, favouring zero-field dynamics over static in-plane states. Above a critical value, zero-field dynamics are no longer observed. The optimum ratio of in-plane shape to out-of-plane uniaxial anisotropy, for which large angle out-of-plane zero-field dynamics occur within the widest current range, is reported.}

\begin{document}
\maketitle

The discovery of giant magnetoresistance effect by Fert\cite{Baibich_1988} and Gr{\"u}nberg\cite{Binasch_1989} in the late 1980's  led to a paradigm shift in the miniaturization of magnetic storage devices which is still visible to the present day. It was not until several years later, that the reciprocal effect, the spin-transfer torque (STT), was theoretically predicted\cite{Slonczewski_1996,Berger_1996}, and several additional years of technological developments in order for this effect to be experimentally observed\cite{Katine_2000,Kiselev_2003}. This initial demonstration spearheaded a rapidly growing field which includes spin-transfer-torque random access memory (STT-MRAM)\cite{Katine_2000}, and spin-torque nano-oscillators (STNOs). STNOs are potential low input power radio-frequency devices for wireless communication, whose frequency can be adjusted simply by changing the applied electrical bias\cite{Kiselev_2003}.

While initial studies on spin-transfer driven dynamics were carried out on fully metallic systems with both the free and the reference layers magnetized in-plane\cite{Kiselev_2003}, hybrid device geometries combining an in-plane (IP) and an out-of-plane (OOP) magnetized layers are presently used\cite{Zeng_2013,Rippard_2010,Kubota_2013,Taniguchi_2013}. This geometry helps to reduce the critical current\cite{Mangin_2006}, maximises the output power due to large angle precession\cite{Zeng_2013,Kubota_2013}, and can provide functionality regardless of applied magnetic or current history\cite{Kubota_2013,Rippard_2010}. Currently, the STNOs based on magnetic tunnel junctions (MTJs) attract the most interest due to their much higher output powers\cite{Kubota_2013} and lower operation currents\cite{Zeng_2013,Deac_2008,Skowronski_2012} compared to their fully metallic counterparts\cite{Kiselev_2003,Rippard_2010}.

In terms of prospective applications, it is desirable that STNOs function without a need for external fields. According to some theoretical models\cite{Taniguchi_2013,Taniguchi_2014_IEEE}, for hybrid geometry MTJs, as considered here, there is no current-driven dynamics at zero applied field.
It has been recently suggested that the perpendicular (field-like) spin-transfer torque may bring about the stabilization of dynamics at zero-field for $\frac{\text{STT}_{\|}}{\text{STT}_{\bot}}\,<\,0$\cite{Taniguchi_2014_APL,Guo_2015}, where STT$_{\|}$ and STT$_{\bot}$ are the in-plane and the perpendicular STT terms, respectively. 
Nevertheless, this has not been experimentally proven and, so far, zero-field dynamics have only been observed in systems having elliptical cross-section\cite{Kiselev_2003,Zeng_2013,Skowronski_2012}. What is lacking is a comprehensive study regarding the influence of shape anisotropy in the considered system.

\begin{figure}[ht!]
\centering
\includegraphics[width=8cm]{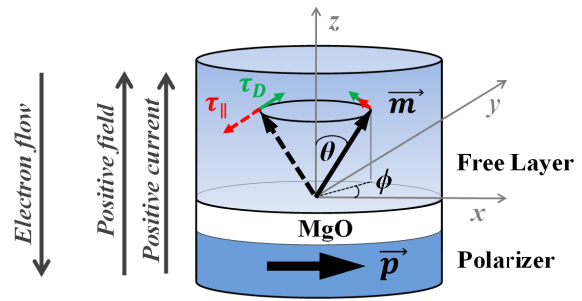}
\caption{Considered STNO geometry with marked directions of the positive field and current. The steady-state precession occurs only for electrons flowing from the polarizer to the free layer\cite{Zeng_2013,Kubota_2013,Skowronski_2012,Taniguchi_2013,Taniguchi_2014_IEEE} (positive current flow), which corresponds to the presented configuration of the in-plane spin-transfer torque and the damping torque (marked as $\tau_{\|}$ and $\tau_D$, respectively).}
\label{fig_1}
\end{figure}

In this paper we analytically and numerically investigate the influence of an in-plane anisotropy component on zero-field out-of-plane dynamics in MgO-based spin-torque nano-oscillators with an OOP magnetized free layer and an IP polarizer (see Fig.~\ref{fig_1}). We vary the magnitude of the in-plane anisotropy of the system by changing the ellipticity of the free layer cross-section, and investigate its influence on the presence of the OOP steady-state precession over given ranges of applied currents and OOP magnetic fields.
We analytically solve the Landau-Lifshitz-Gilbert-Slonczewski (LLGS) equation for a typical device with circular cross-section under perpendicular applied fields and currents. We define the angular asymmetry of the in-plane term of the spin-transfer torque as resulting from the cosine-type dependence of the tunnel magnetoresistance (TMR) on the angle between the magnetizations of the free and the reference layers\cite{Slonczewski_1989,Moodera_1996} (i.e., {\bf m} and {\bf p} vectors, respectively; see Fig.~\ref{fig_1}). 
We take into account the spin-torque bias dependence, as well as the bias dependence of the tunnel magnetoresistance, which, in fact, yields a suppression of the output power at large applied currents (as demonstrated in Ref.[19]). We assume a linear bias dependence of the TMR for the antiparallel (AP) state ($R_{AP}$) and a constant resistance for the parallel (P) state ($R_P$)\cite{Slonczewski_2005,Kalitsov_2013,Moodera_1996,Gao_2007,Kowalska}.

The motion of the free layer magnetization ${\bf m}$ is described by the LLGS equation\cite{Slonczewski_1996,Deac_2008}:

\begin{equation}
\frac{d{\bf m}}{dt} = - \gamma ({\bf m} \times {\bf B_{eff}}) + \alpha ({\bf m} \times \frac{d{\bf m}}{dt}) + \gamma \frac{\partial \tau_{\|}}{\partial V} V [ {\bf m} \times ({\bf m} \times {\bf n_x}) ].
\label{eq_1}
\end{equation}

Here, $\gamma$ is the gyromagnetic ratio, $\frac{\partial \tau_{\|}}{\partial V} [\frac{T}{V}]$ is the torkance of the in-plane STT term\cite{Kubota_2008,Theodonis_2006}, $\alpha$ is a Gilbert damping constant, {$\bf n_x$} and {$\bf n_z$} are the unit vectors of the coordinate system presented in Fig.~\ref{fig_1}. The effective magnetic field is defined as ${\bf B_{eff}} = B_{ext} {\bf n_z} + B_{k_{\bot}} m_z {\bf n_z} + B_{k_{\|}} m_x{\bf n_x}$, where $B_{ext}$ is the external field, $B_{k_{\bot}}$ is the effective out-of-plane anisotropy along the $z$-axis ($B_{k_{\bot}} = B_k - \mu_0 M_S$, where $B_k$ is an uniaxial magnetic anisotropy field and $M_s$ is a saturation magnetization), and $B_{k_{\|}}$ is the effective in-plane shape anisotropy along the $x$-axis.

We define a linear bias dependence of the resistance difference $\Delta R$ between the P and the AP states as follows: $\Delta R = - \frac{\partial R_{\text{AP}}}{\partial V} \cdot |V| + \Delta R_0$ (here, $\Delta R_0$ is the resistance difference between the two states close to zero bias, and $\frac{\partial R_{\text{AP}}}{\partial V}$ is the slope of the linear bias dependence of the AP state resistance)~\cite{Kowalska}. For each instant angle between the magnetic moments of two layers, we convert the current $i$ into an equivalent voltage value $V$, as we have previously proposed in Ref.[19].
In order to estimate the onset current for precession, we consider the limit of small precession angles $\theta$ (i.e., $\theta \rightarrow 0$ for positive applied fields and $\theta \rightarrow \pi$ for negative applied fields), and we solve the LLGS equation (\ref{eq_1}) expressed in the spherical coordinates, as discussed in detail in Ref.[19]. We obtained the following analytical equation defining the region of the out-of-plane steady-state precession:

\begin{equation}
B_{ext (\theta \rightarrow 0)} (i) < \Big| \frac{ \frac{\partial \tau_{\|}}{\partial V} \big( \Delta R_0 - |i| \frac{\partial R_{AP}}{\partial V} R_P \big)}{\alpha \big( 2 + |i| \frac{\partial R_{AP}}{\partial V} \big)^2} i - B_{k_{\bot}} + B_{k_{\|}} \Big|.
\label{eq_14}
\end{equation}

For $B_{k_{\|}} = 0$, at low applied fields, the in-plane STT stabilizes the static in-plane antiparallel state within a region of the phase diagram defined as follows\cite{Kowalska}:

\begin{equation}
B_{ext}(i) \leq \big| \frac{\partial \tau_{\|}}{\partial V} R_P i \big|.
\label{eq_x}
\end{equation}

In order to prove the validity of the analytical solutions, the numerical integration of eq.~\ref{eq_1} was also performed. We used the MAPLE 8 program and the simulation parameters typical for the considered system: $\frac{\partial \tau_{\|}}{\partial V}$\,=\,0.00672\,$\frac{T}{V}$\cite{Jung_2010}, $\alpha=$\,0.005\cite{Okada_2014}, $\Delta R_0=$\,100\,$\Omega$\cite{Kowalska}, $R_P=$\,200\,$\Omega$\cite{Kowalska}, $B_{k_{\bot}}=$\,200\,$mT$, and $\frac{\partial R_{AP}}{\partial V}=$\,10\,$\frac{\Omega}{V}$. The simulation time was set to 150\,ns and the initial magnetization position was random. The final static and dynamic states were defined based on the last 2\,ns of the simulation.

\begin{figure}[H]
\centering
\includegraphics[width=8.8cm]{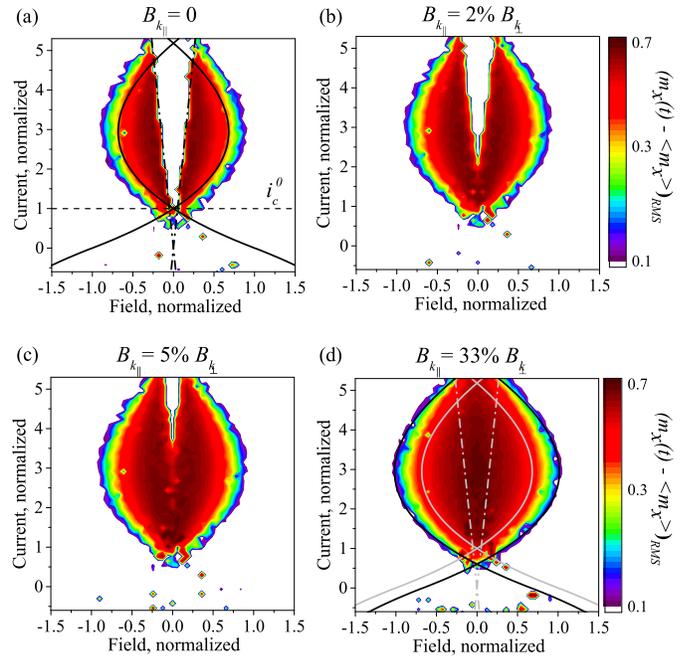}
\caption{Current versus field dynamic state diagrams: (a) for a nano-pillar with circular cross-section ($B_{k_{\|}} = 0$), and for one with elliptical cross-section where (b) $B_{k_{\|}}\,=\,2\% B_{k_{\bot}}$, (c) $B_{k_{\|}}\,=\,5\% B_{k_{\bot}}$, (d) $B_{k_{\|}}\,=\,33\% B_{k_{\bot}}$. Coloured area shows numerically determined intensity of magnetization dynamics along $x$-axis. Black solid lines in (a) and (d) show analytical solution determining the region of steady state precession (eq.~\ref{eq_14}). Dashed lines in (a) determine the stability region of static IP state for the case of $B_{k_{\|}} = 0$ (eq.~\ref{eq_x}). For the comparison, grey solid and dashed lines in (d) show the analytical solution for the case shown in (a).}
\label{fig_2_s}
\end{figure}

Fig.~\ref{fig_2_s} shows numerically determined field versus current dynamic phase diagrams calculated for free layers with different cross-sectional ellipticities. The coloured areas show the intensity of magnetization dynamics along the $x$-axis, expressed as the root mean square (RMS) of $(m_x(t) - \langle m_x \rangle)$ (where $m_x(t)$ and $\langle m_x \rangle$ are the instantaneous and the mean $m_x$ value, respectively). In a real device, these values directly correlate with the generated output power of the spin-torque nano-oscillator.\cite{Fowley_2014}
Normalized current values are defined as $\frac{i}{i_{c}^{0}}$, where $i_{c}^{0}$ is the current value at the crossing of the analytically determined critical lines for dynamics for the case of a nano-pillar with circular cross-section (as marked with dashed line in Fig.~\ref{fig_2_s}(a)). Field values are normalized by the effective out-of-plane anisotropy, $B_{k_{\bot}}$.

The dynamic diagram for the case of the circular cross-section, i.e. when the in-plane anisotropy component is equal to zero ($B_{k_{\|}} = 0$), is presented in Fig.~\ref{fig_2_s}(a). Our results show that stable dynamics occur only for positive currents, defined as electrons flowing from the free to the reference layer\cite{Zeng_2013,Kubota_2013,Skowronski_2012,Taniguchi_2013,Taniguchi_2014_IEEE}. This is a consequence of the fact that, in most MTJs, STT$_{\|}$ is larger close to the AP state and, thus, more capable of overcoming the damping torque\cite{Kowalska} (see a sketch of precession mechanism in Fig.~\ref{fig_1}). As shown in Fig.~\ref{fig_2_s}(a), with increasing applied field, the onset current for dynamics initially increases quasi-parabolically. Above a certain current value (here, for $i=$\,3\,$\cdot i_{c}^{0}$), we observe a quenching of STNO dynamics and a reduction of the dynamic region, which is a direct effect of the TMR bias dependence\cite{Kowalska}.
We also observe a gap in the dynamics at zero and low applied field, where the antiparallel static state is stabilized\cite{Kowalska}.
The analytical solution defining the area of out-of-plane dynamics (eq.~\ref{eq_14}) and the static in-plane state (eq.~\ref{eq_x}) are plotted in Fig.~\ref{fig_2_s}(a) with solid and dashed lines respectively, and accurately define the regions of numerically obtained out-of-plane dynamics.

We now define the ellipticity of the free layer by introducing the in-plane shape anisotropy, $B_{k_{\|}}$, expressed in the percentage of the effective out-of-plane anisotropy, $B_{k_{\bot}}$.
While increasing $B_{k_{\|}}$ from 2\% (Fig.~\ref{fig_2_s}(b)) to 5\% (Fig.~\ref{fig_2_s}(c)) of $B_{k_{\bot}}$, the dynamic gap from Fig.~\ref{fig_2_s}(a) gradually closes and zero-field dynamics are stabilized over the entire current range for $B_{k_{\|}}=$\,10\% $B_{k_{\bot}}$.
This is the effect of the gradual tilting of the magnetization towards the in-plane easy direction ($x$-axis) which enables the onset angle for precession to be reached at lower electrical currents, leading to stable magnetization dynamics and a suppression of the critical behaviour of the stabilization of static AP state.
Further increase of $B_{k_{\|}}$ leads to the largest intensity of zero-field magnetization dynamics at $B_{k_{\|}}=$\,33\% $B_{k_{\bot}}$ (see Fig.~\ref{fig_2_s}(d)). We also observe here a pronounced expansion of the dynamic region along the field axis, which can be easily distinguished by comparing the plotted analytical lines of these two cases (i.e., the black and the grey lines in Fig.~\ref{fig_2_s}(d), showing the analytical solutions for $B_{k_{\|}}=$\,33\% $B_{k_{\bot}}$ and $B_{k_{\|}}=$\,0, respectively).

\begin{figure}[H]
\centering
\includegraphics[width=8.6cm]{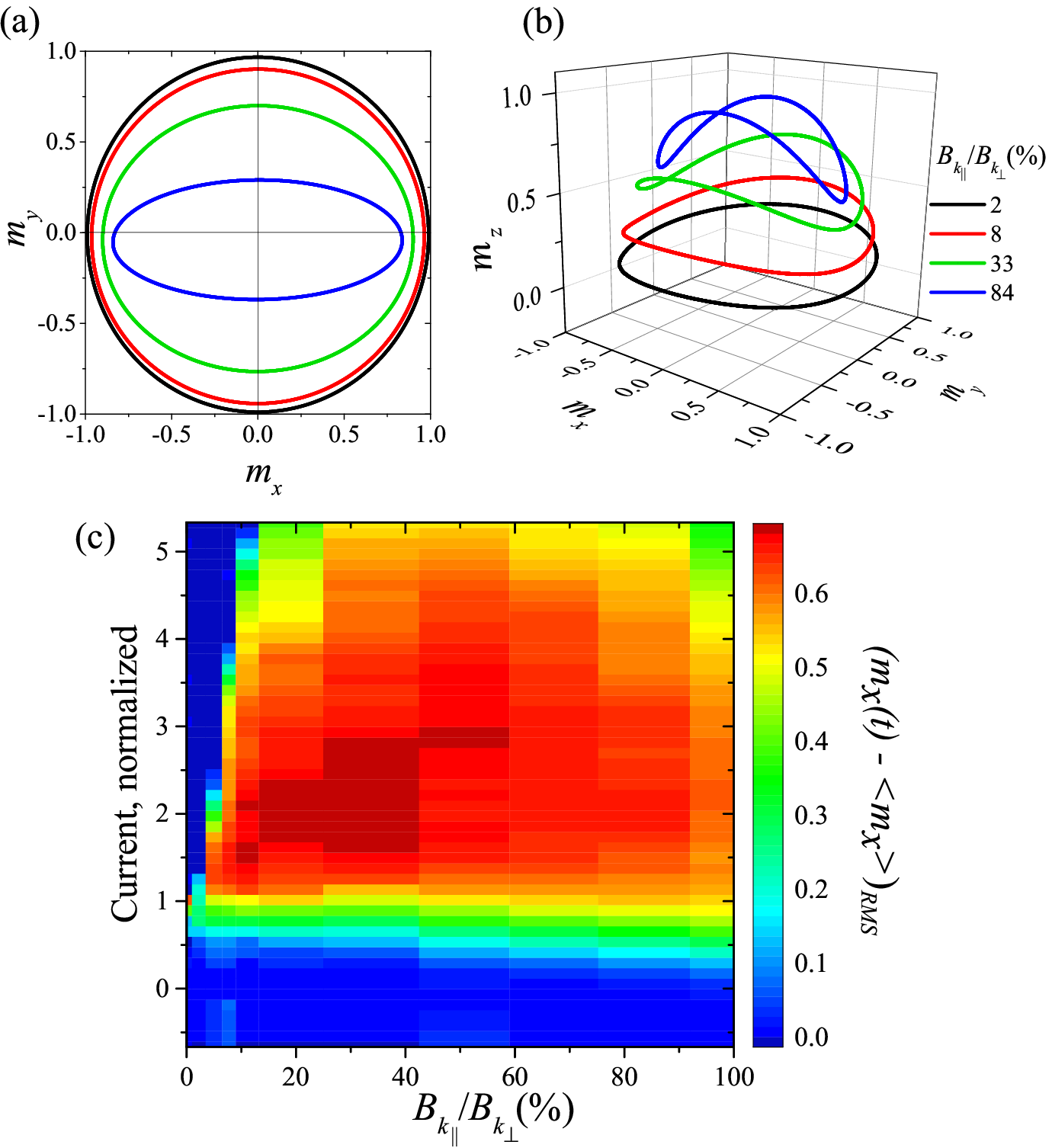}
\caption{Zero-field magnetization dynamics for $0 < B_{k_{\|}} < B_{k_{\bot}}$. Magnetization precession trajectories at $i = 1.9 \cdot i_{c}^{0}$ for different $B_{k_{\|}}$ values: (a)~projections of trajectories on $xy$-plane, (b) trajectories in $xyz$ space. (c), Intensity of magnetization dynamics as a function of the applied current and $B_{k_{\|}}/B_{k_{\bot}}$ expressed in percentage.}
\label{fig_6}
\end{figure}

The comparison between the shapes of the magnetization precession trajectories for different $B_{k_{\|}}$ values is presented in Fig.~\ref{fig_6}(a) and (b). Fig.~\ref{fig_6}(a) shows concentric projections of the precession trajectories on the $xy$-plane, where the ellipticity increases with the increasing $B_{k_{\|}}$. We can see a gradual decrease of the maximum $m_x$ and $m_y$ components, with a more pronounced reduction of $m_y$. 3-dimensional plots in Fig.~\ref{fig_6}(b) shows the real shapes of the trajectories, which are transforming from circular (black trajectory) to out-of-plane clam-shell-type (blue trajectory) while increasing $B_{k_{\|}}$ from $B_{k_{\|}}=$\,2\% $B_{k_{\bot}}$ to $B_{k_{\|}}=$\,84\% $B_{k_{\bot}}$. This directly shows how the anisotropy energy, arising from the introduced elliptical shape of the nano-pillar, acts on the magnetization by pushing it away from the in-plane magnetic hard axis, oriented along the $y$-axis, similar to what is obtained in equivalent metallic systems\cite{Fowley_2014}.

Fig.~\ref{fig_6}(c) shows the intensity of zero-field magnetization dynamics along the $x$-axis as a function of $B_{k_{\|}}/B_{k_{\bot}}$ (in \%) and the current. The minimum current required to observe dynamics is reduced gradually by around $\frac{i}{i_{c}^{0}}=$\,0.3 as $B_{k_{\|}}$ is increased from $B_{k_{\|}}$\,=\,0 to $B_{k_{\|}}=$\,100\% $B_{k_{\bot}}$.
Maximum operation currents (i.e., high currents still driving dynamics) initially increase with increasing $B_{k_{\|}}$, up to the $B_{k_{\|}}=$\,50\% $B_{k_{\bot}}$, and then decreases.
We also observe a general increase of the intensity of magnetization dynamics up to $B_{k_{\|}}=$\,35\% $B_{k_{\bot}}$, followed by a decrease as $B_{k_{\|}}$ further increases.
According to Fig.~\ref{fig_6}(c), the in-plane anisotropy should be then optimized for both low onset currents and large operation current ranges, simultaneously still maintaining high output power zero-field dynamics of the device.
We found the optimum value of $B_{k_{\|}}=$\,33\% $B_{k_{\bot}}$, where the large intensity of zero-field magnetization dynamics were observed within the widest current range and the maximum $(m_x(t) - \langle m_x \rangle)_{RMS}\approx$\,0.7 was obtained for the largest current range of $1.5 < \frac{i}{i_{c}^{0}} < 3$). Note that the dynamic diagram for $B_{k_{\|}}=$\,33\% $B_{k_{\bot}}$ is also presented in Fig.~\ref{fig_2_s}(d).

\begin{figure*}[ht]
\centering
\includegraphics[width=\textwidth]{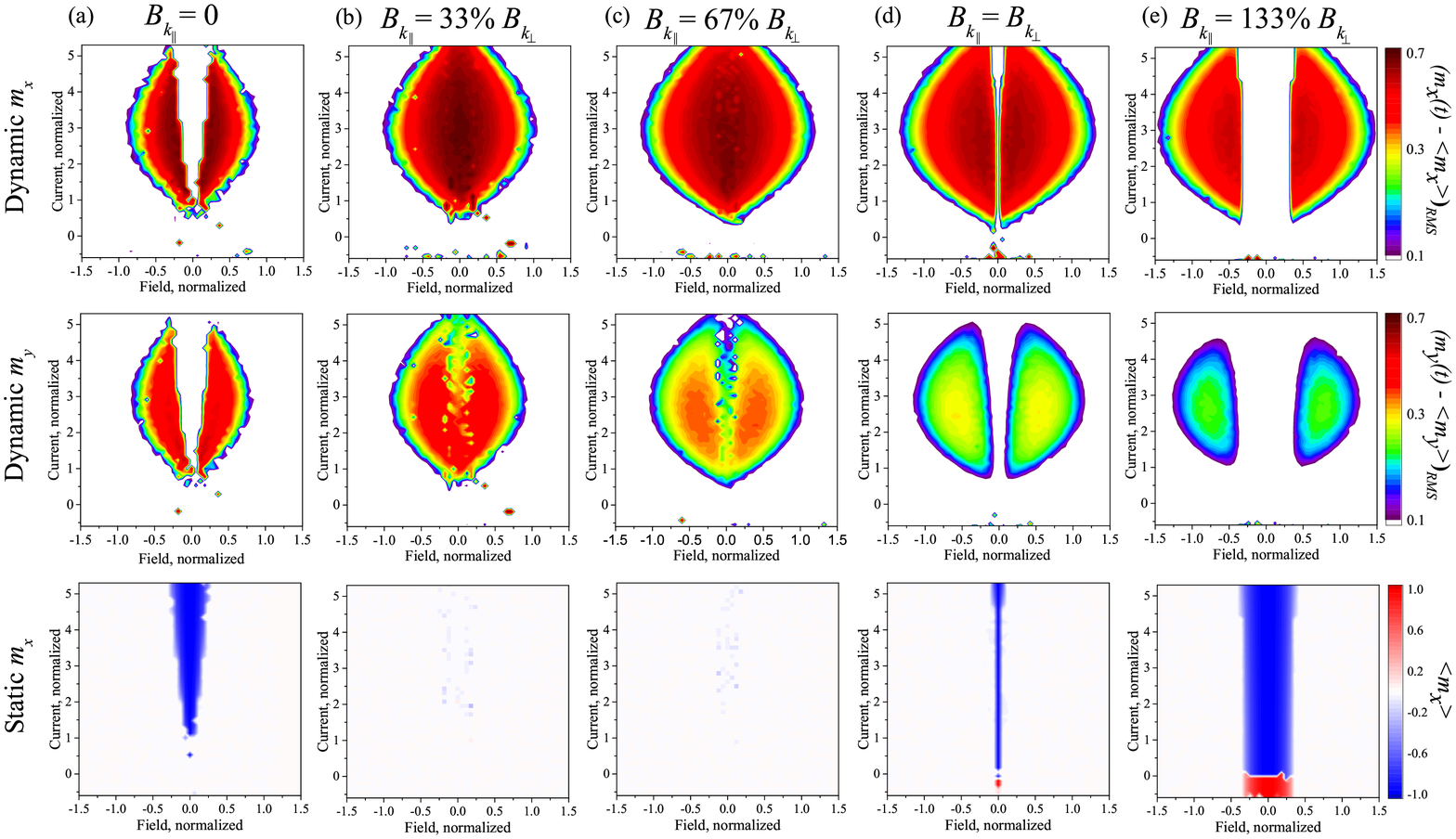}
\caption{Dynamic and static phase diagrams for: (a) $B_{k_{\|}}=$\,0, (b) $B_{k_{\|}}=$\,33\% $B_{k_{\bot}}$, (c) $B_{k_{\|}}=$\,67\% $B_{k_{\bot}}$, (d) $B_{k_{\|}} = B_{k_{\bot}}$, and (e) $B_{k_{\|}}=$\,133\% $B_{k_{\bot}}$. First row: corresponding phase diagrams showing the intensity of magnetization dynamics along the $x$-axis, translating directly into the output power of the STNO device. Second row: corresponding phase diagrams of the intensity of magnetization dynamics along the $y$-axis, reflecting a progressing deformation of the magnetization precession trajectory with increasing $B_{k_{\|}}$. Third row: corresponding phase diagrams of the static in-plane state, showing an average magnetization along the $x$-axis; here, the region of $\langle m_x \rangle=$\,-1 represents the static in-plane AP state.}
\label{fig_3_s}
\end{figure*}

Fig.~\ref{fig_3_s} shows the effect of the increasing in-plane shape anisotropy, $B_{k_{\|}}$, on the out-of-plane dynamics along the $x$-axis, $(m_x(t) - \langle m_x \rangle)_{RMS}$ (the first row), and along the $y$-axis, $(m_y(t) - \langle m_y \rangle)_{RMS}$ (the second row), as well as on the stability region of static in-plane states, represented by an average $m_x$ component, $\langle m_x \rangle$ (the third row).
With increasing $B_{k_{\|}}$, we observe a general reduction of intensity of magnetization dynamics along the $x$-axis (see the first row), translating into the loss of the output power in the actual STNO device.
This is consistent with the evolution of the shape of the magnetization trajectory presented above, which shows that the maximum value of $m_x$ decreases (see trajectory projections in Fig.~\ref{fig_6}(a)) once the trajectory changes its shape from circular to clam-shell-type (as shown in Fig.~\ref{fig_6}(b)).
This also explains the significant reduction of the intensity of magnetization dynamics along the $y$-axis (see the second row in Fig.~\ref{fig_3_s}), occurring due to the described deformation of the precession trajectory.

The third row in Fig.~\ref{fig_3_s} shows the corresponding average magnetization along the $x$-axis. Here, the blue areas represent the stability regions of the static AP state (for $\langle m_x \rangle=$\,-1). For a free layer with circular cross-section, shown in Fig.~\ref{fig_3_s}(a), at small applied fields, positive currents stabilize the static in-plane AP state. For the cases of $B_{k_{\|}}=$\,33\% $B_{k_{\bot}}$ and $B_{k_{\|}}=$\,67\% $B_{k_{\bot}}$ (the third row of Fig.~\ref{fig_3_s} (b) and (c)), the in-plane anisotropy stabilizes zero-field dynamics, replacing the area of the static AP state. 
When $B_{k_{\|}} = B_{k_{\bot}}$ (Fig.~\ref{fig_3_s}(d)) or $B_{k_{\|}} > B_{k_{\bot}}$ (Fig.~\ref{fig_3_s}(e)), no dynamics are obtained at zero-field (neither along $x$- nor $y$-axis), and the static AP state at $\langle m_x \rangle=$\,-1 is instead stabilized. We additionally observe the~presence of an onset field for out-of-plane dynamics which is, in fact, slightly larger than the difference between the two anisotropies, i.e. $B_{onset} \approx B_{k_{\|}}-B_{k_{\bot}}$. As a result, in the case when $B_{k_{\|}} > B_{k_{\bot}}$, one should apply an external out-of-plane field in order to overcome the in-plane anisotropy and pull the magnetization back towards the OOP direction.

To summarize, we numerically and analytically investigate the influence of the in-plane shape anisotropy on zero-field out-of-plane dynamics in MgO-based STNOs with an OOP magnetized free layer and an IP polarizer. 
As previously reported, we observe no zero-field dynamics for circular nano-pillars of this particular geometry. 
According to our results, introducing a $B_{k_{\|}}$ constant as low as 2\% of $B_{k_{\bot}}$ is already sufficient for inducing zero-field oscillations, which persist as long as $B_{k_{\|}} < B_{k_{\bot}}$. Initially, the oscillations are restricted to a narrow current range, which increases as $B_{k_{\|}}$ increases, covering the whole dynamic range for $B_{k_{\|}}=$\,10\% $B_{k_{\bot}}$. 
When $B_{k_{\|}} \geq B_{k_{\bot}}$, the anisotropy stabilizes the static in-plane AP state at zero-field, which prevents precession. Dynamics can be recovered by the application of a critical field $B_{onset}$ which is approximately equal to $B_{k_{\|}}-B_{k_{\bot}}$. We also observed a general decrease of the intensity of magnetization dynamics with increasing ellipticity, which is assigned to a gradual change of the magnetization precession trajectory from the circular to clam-shell-type.
Consequently, the shape of STNO nano-pillars should be taken into account while designing an actual commercial device; in particular, in terms of the presence of zero-field dynamics itself, but also in order to maximize the output power, minimize the operation currents, or tune the operation current range.

\acknowledgment
This research was supported by the Helmholtz Young Investigator Initiative Grant No. VH-N6-1048. The authors would like to thank Dr. Kai Wagner for fruitful discussions.

\end{document}